\documentclass[12pt,reqno]{article}

\usepackage{nopageno}

\usepackage{setspace}
\usepackage{latexsym}
\usepackage{amsthm}
\usepackage{amssymb}
\usepackage{amsmath}
\usepackage{amsfonts,amsgen,amsopn}
\usepackage{authblk}

\usepackage{epsfig}
\usepackage{color}
\usepackage[font=scriptsize,labelfont=bf]{caption}
\usepackage{subfig}
\usepackage[a4paper, hdivide={2.5cm,,2.5cm}, vdivide={3.5cm,,2cm}]{geometry} 
\usepackage{natbib}
  
\usepackage[english]{babel}
\usepackage{fancyheadings}
\newtheorem{theorem}{Theorem}[section]
\newtheorem{lemma}[theorem]{Lemma}

\newtheorem{corollary}[theorem]{Corollary}
\newtheorem{proposition}[theorem]{Proposition}

\newtheorem*{acknowledgement}{Acknowledgement}

\newtheorem{conjecture}{Conjecture}
\theoremstyle{remark}
\newtheorem*{idea}{\bf Main Idea}

\newcommand{\T}{{\mathcal T}}
\newcommand{\C}{{\mathcal C}}

\newcommand{\NN}{{\mathbb N}}

\begin{document}

\bibliographystyle{plainnat}

\begin{titlepage}
  \title{\bfseries Non-hereditary maximum parsimony trees}
  \pagestyle{headings} \markboth{Non-hereditary MP trees}{Non-hereditary MP trees} \date{}

  \author{\normalsize Mareike Fischer} 
  \affil{\footnotesize Center for Integrative Bioinformatics (CIBIV), Dr. Bohr Gasse 9, 1030 Vienna, Austria;  \\E-mail: email@mareikefischer.de}

\end{titlepage}
\maketitle

\begin{abstract}
   In this paper, we investigate a conjecture by von Haeseler concerning the Maximum Parsimony method for phylogenetic estimation, which was published by the Newton Institute in Cambridge on a list of open phylogenetic problems in 2007. This conjecture deals with the question whether Maximum Parsimony trees are hereditary. The conjecture suggests that a Maximum Parsimony tree for a particular (DNA) alignment necessarily has subtrees of all possible sizes which are most parsimonious for the corresponding subalignments. We answer the conjecture affirmatively for binary alignments on five taxa but also show how to construct examples for which Maximum Parsimony trees are not hereditary. Apart from showing that a most parsimonious tree cannot generally be reduced to a most parsimonious tree on fewer taxa, we also show that compatible most parsimonious quartets do not have to provide a most parsimonious supertree. Last, we show that our results can be generalized to Maximum Likelihood for certain nucleotide substitution models.

  \par
  \vspace{0.3cm} {\noindent \bfseries Keywords:} phylogenetics, maximum
  parsimony, maximum likelihood, Jukes-Cantor model
\end{abstract}

\newpage \setcounter{page}{1}

\section{Introduction}\label{introduction}
Tree reconstruction methods for inferring phylogenetic trees are used to interpret
the ever-growing amount of available genetic sequence
data. Unsurprisingly, such methods have therefore been widely
discussed in the last decades (e.g., \citep{felsenstein_1978};
\citep{felsenstein_2004}; \citep{semple_steel_2003};
\citep{yang_2006}). One of the most frequently used tree reconstruction
methods is the so-called Fitch parsimony \citep{fitch_1971} or {\em Maximum Parsimony} method (MP). Two of the reasons for its popularity are its simplicity compared to other methods such as Maximum Likelihood as well as its purely combinatorial basic principle. The latter makes MP a method that can be applied to any data alignment without any assumptions on the way the data has been generated, which means for the DNA that no assumptions on the probability of a nucleotide substitution have to be made (which is why MP is often said to be `model-free'). Despite this simplicity, not all aspects of MP are to-date understood. One of the questions that remained unsolved for quite some time is whether MP trees are hereditary, i.e. if for an MP tree of an alignment on $m$ taxa we can find a subtree of this tree of size $k$ (for all $k=4,\ldots, m-1$) which is most parsimonious for the corresponding subalignment. This problem was submitted by Arndt von Haeseler to the Isaac Newton Institute's list of open phylogenetic problems in 2007 (see http://www.newton.ac.uk/programmes/PLG/conj.pdf) as well as to the `Penny Ante' list of the Annual New Zealand Phylogenetics Meeting in Kaikoura in 2009 (see http://www.math.canterbury.ac.nz/bio/events/kaikoura09/penny.shtml). 
The importance of the conjecture is manifold. Biologically, MP trees with no MP subtrees seem quite counterintuitive as one would expect the MP tree to be related to MP trees on fewer taxa. Particularly when outgroups are included in a DNA analysis, one would want the topology of the rest of the tree to be independent of the outgroup, so the topology of the `best' tree should be independent of the presence or absence of the outgroup. Moreover, if the conjecture was true, there would be sequences of MP trees, starting from four taxa and growing one new leaf at a time, leading to each MP tree of the whole alignment under consideration, so the big trees would `inherit' their MP property from smaller trees. Mathematically, such a property would be particularly interesting with regards to inductive proofs or dynamic programming. However, we  show in this paper that the conjecture is not in general true, but does hold in some special cases like in the case when the alignment is homoplasy-free or when the alignment is binary and there are only five taxa. 

While the above mentioned aspects of heredity basically refer to reducing large MP trees to smaller ones, we also consider the opposite scenario: we show that even if an alignment has only unique MP quartet trees for all 4-taxa subalignments and even if all these quartets are compatible with one another, the supertree comprising all these quartets is not necessarily an MP tree for the original alignment. This means that MP quartets cannot generally be combined into larger MP trees.

Last, we investigate the impact these findings concerning MP have on Maximum Likelihood (ML) under the (generalized) Jukes-Cantor model (also known as $N_r$-model). In this analysis, we use the strong relationship of MP and ML as described in \citep{tuffley_steel_1997} and conclude that the cases that are problematic for MP also turn out to be problematic for ML under the $N_r$-model, even if there is a common mechanism of site evolution.

\section{Notation and Model Assumptions}
\label{notation}

Recall that an {\it unrooted binary phylogenetic $X$-tree} is a tree $\T =(V(\T),E(\T))$
on a leaf set $X=\{1,\ldots,m\} \subset V(\T)$ with only vertices of degree 1 (leaves) or 3 (internal vertices). In this paper, when there is no ambiguity we often just write `tree' when referring to an unrooted binary phylogenetic $X$-tree. Furthermore,
recall that a {\it character} $f$ is a function $f: X\rightarrow \C$ for
some set $\C:=\{c_1, c_2, c_3, \ldots, c_r \}$ of $r$ {\em character
  states} ($r \in \NN$). An {\it extension} of $f$ to $V(\T)$ is a map
$g: V(\T)\rightarrow \C$ such that $g(i) = f(i)$ for all $i$ in $X$. For
such an extension $g$ of $f$, we denote by $l_{\T}(g)$ the number of
edges $e=\{u,v\}$ in $\T$ on which a substitution occurs,
i.e. where $g(u) \neq g(v)$. The {\em parsimony score} of $f$ on $\T$,
denoted by $l_{\T}(f)$, is obtained by minimizing $l_{\T}(g)$ over all
possible extensions $g$. The parsimony score of a sequence of characters $S:= f_1f_2\ldots f_n$ is given by $l_{\T}(S)=\sum\limits_{i=1}^n\;l_{\T}(f_i)$. Note that $S$ cannot only be viewed columnwise as a sequence of characters, but also rowwise as aligned (DNA) species data. In this paper, we therefore use the terms `sequence of characters' and `alignment' synonymously when there is no ambiguity. Moreover, we denote by $f-k$, $S-k$ and $\T-k$ the restriction of $f$, $S$ and $\T$, respectively, on the set $X-k$; so $k \in X$ is the taxon that is present in $f$, $S$ and $\T$ but not in $f-k$, $S-k$ and $\T-k$.

A character $f$ is said to be {\it homoplasy-free} on a tree $\T$ if $l_{\T}(f)=|f|-1$ , where $|f|$ denotes the number of character states employed by $f$. A sequence $S$ of characters is called homoplasy-free when all its characters have that property. Note that if a character or an alignment is homoplasy-free on a certain tree, this tree minimizes its parsimony score and is therefore most parsimonious for this character or alignment, respectively.

Recall that a character $f$ on a leaf set $X$ is said to be {\em informative}
(with respect to parsimony) if at least two distinct character states
occur more than once on $X$. Otherwise $f$ is called {\em non-informative}. Note
that for a non-informative character $f$, $l_{\T_i}(f)=l_{\T_j}(f)$ for
all trees $\T_i$, $\T_j$ on the same set $X$ of leaves. In this paper, we refer to a character always with its underlying taxon clustering pattern in mind, i.e. for instance we do not distinguish between $AACC$, $CCAA$ and $CCGG$, and so on.
  
Next we describe the fully symmetric $r$-state model
\citep{neyman_1971}, also known as the $N_r$-model, which underlies the
Tuffley and Steel equivalence result concerning MP and ML \citep{tuffley_steel_1997}.

Consider a phylogenetic $X$-tree $\T$ arbitrarily rooted at one of its
vertices. The $N_r$-model assumes that a state is assigned to the root
from the uniform distribution on the set of states. The state then
evolves away from the root as follows. The model assumes equal rates of
substitutions between any two distinct character states. For any edge $e
= \{u,v\} \in E(\T)$, where $u$ is the vertex closer to the root, let
$p_e$ denote the conditional probability $P(v=c_i|u=c_j)$, where $c_i
\neq c_j$. The probability $p_e$ is equal for all pairs of distinct
states $c_i$ and $c_j$. Therefore, the probability that a substitution
($c_j$ to a state different from $c_j$) occurs on the edge $e$ is
$(r-1)p_e$. Let $q_e$ be the conditional probability $P(v=c_i|u=c_i)$,
i.e. the probability that no substitution occurs on edge $e$. In the
$N_r$-model, we have $0\leq p_e\leq \frac{1}{r}$ for all $e \in E(\T)$,
and $(r-1)p_e + q_e = 1$. Moreover, the $N_r$-model assumes that
substitutions on different edges are independent. Note that for $r=4$,
the $N_r$-model coincides with the well-known Jukes-Cantor model
\citep{jukes_cantor_1969}.

Let $\T$ be a phylogenetic $X$-tree and let $f$ be a character on its
leaf set $X$.  Let the substitution probabilities assigned to the edges
of $\T$ under the $N_r$-model be collectively denoted by $\bar{p}:=
(p_e: e\in E(\mathcal{T}))$. Then we denote by $P(f|\T, \bar{p})$ the
probability of observing character $f$ given tree $\T$ and the
parameter values $\bar{p}$. Note that $P(f|\T, \bar{p})$ does not depend
on the root position since the model is symmetric. The maximum value of
this probability for fixed $f$ and $\T$ as $\bar{p}$ ranges over all
possibilities is denoted by $\max P(f|\T)$, i.e. $\max P(f|\T) :=
\max_{\bar{p}}P(f|\T,\bar{p})$.

Now let $S:=f_1\ldots f_n$ be a sequence of characters. When we refer to a sequence of characters under the $N_r$-model with {\em no common mechanism}, this means that the substitution probabilities on
edges may be different for different characters in $S$ without any
correlation between the characters. We suppose that for each character
$f_i$ in the sequence and for each edge $e$ of the tree, there is a
parameter $p_{e,i}$ that gives the substitution probability for $f_i$ on
edge $e$. When there is no common mechanism, the parameters $p_{e,i}$ are all independent. 
For $i=1,\ldots,n$, let $\bar{p}_i:= (p_{e,i}: e\in E(\mathcal{T}))$ be the
vectors of substitution probabilities. We denote the model parameters
$(\bar{p}_i, i=1,\ldots,n)$ collectively as $\Theta$ and refer to $P(S|\T, \Theta)$ as the probability of observing sequence $S$
given the phylogenetic tree $\T$ and model parameters $\Theta $. We then
define the likelihood of the tree $\T$ and the model parameters $\Theta $
given the sequence $S$, which we refer to as the {\em likelihood function}, as
$L(\T, \Theta|S) := P(S|\T, \Theta)$. The maximum likelihood method of
phylogenetic tree reconstruction involves optimizing the likelihood
function in two steps as described in \citep{semple_steel_2003}. We
first maximize $P(S|\T,\Theta)$ over the space of model parameters
$\Theta$. We define:
\begin{equation*}
  \max P(S|\T) :=  \max_{\Theta }P(S|\T,\Theta).
\end{equation*} 
We then choose a tree $\T$ that maximizes $\max P(S|\T)$. We call such a
tree a maximum likelihood tree (ML-tree) of $S$. Thus, an ML-tree of a
sequence $S$ is $\mathrm{argmax}_{\T}\left( \max
  P(S|\T)\right)$. Note that under the assumption of no common
mechanism, i.e. the characters in an alignment are regarded independent of one another, we have:
\[ \max P(S|\T) =
\prod_{i=1}^n\max_{\bar{p}_i}P(f_i|\T,\bar{p}_i).  \]

\section{Results}

\subsection{Heredity Part I: Inferring small MP trees from larger ones}\label{hereditary1}

As explained in Section \ref{introduction}, we analyze cases in which MP trees are or are not hereditary. In particular, we examine whether a most parsimonious tree on an alignment needs to be related to most parsimonious trees on subalignments. In fact, the conjecture under investigation suggests a sequence of MP trees of sizes leading from the number of taxa considered down to four, where small MP trees are subtrees of the larger ones. Note that as there is only one unrooted tree on one, two and three taxa, respectively, the conjecture does not consider these cases as subtrees of these sizes are unique and therefore always most parsimonious.

We now formulate the conjecture mathematically.

\begin{conjecture} [Conjecture PC5 from the Isaac Newton Institute's `Phylogenetics: Challenges and Conjectures' list 2007] \label{conj} Let $S:= f_1f_2\ldots f_n$ be a sequence of characters (`alignment') on the set $X$ of taxa, where $|X|=m$, and let $\T$ be a Maximum Parsimony tree for $S$. Then, for each $k=4,\ldots, m-1$, there exists a subset $Y$ of $X$ of size $k$ so that $\T|_Y$ is an MP tree for $S|_Y$ (where $S|_Y$ is the sequence $S$ of characters restricted to the taxa in $Y$ and $\T|_Y$ is the tree $\T$ restricted to the taxa in $Y$).
\end{conjecture}

While finding an MP tree is generally NP-hard \citep{foulds_graham_1982}, if this conjecture was true it might be relevant for dynamic programming approaches concerning certain instances of parsimony. Note that the conjecture does not state which particular subtree would be most parsimonious -- so the conjecture is not in conflict with the NP-hardness of Maximum Parsimony and therefore could be valid. Moreover, mathematically a statement like that given in the conjecture would be useful to investigate theoretical properties of MP using inductive proofs, as the inductive step in such proofs requires knowledge on smaller instances of the problem under investigation.

In the following, we will present two special cases of the conjecture, namely the case in which the given alignment is homoplasy-free as well as the case where the alignment is on five taxa and employs only binary characters. In these cases, the conjecture is true. Moreover, we afterwards analyze more general cases where the conjecture fails.

We need the following lemma in order to prove a first positive result concerning Conjecture \ref{conj}.

\begin{lemma} \label{homosubcharacter} Let $\T$ be an unrooted binary phylogenetic $X$-tree for a taxon set $X$ with $|X|=m$ and let $f$ be a homoplasy-free character on $\T$. Let $k \in X$ be a taxon. Then, $f-k$ is homoplasy-free on $\T-k$.  
\end{lemma}

\begin{proof} By definition of homoplasy, $l_{\T}(f)=|f|-1$. For any taxon $k \in X$, note that the parsimony score of the character $f-k$ on $\T-k$ can be calculated as follows: \begin{eqnarray*}l_{\T-k}(f-k)&=&\begin{cases} l_{\T}(f)=|f|-1 & \text{if the character state $c_k$ of taxon $k$ is not unique in $f$,} \\ l_{\T}(f)-1=|f|-2 & \text{else,}\end{cases}\\ & =&\begin{cases} |f-k|-1 & \text{if $|f-k|=|f|,$} \\ |f-k|-1 &  \text{if $|f-k|=|f|-1.$}\end{cases}\end{eqnarray*} So altogether $l_{\T-k}(f-k)= |f-k|-1.$ Thus, $f-k$ is homoplasy-free on $\T-k$.  
\end{proof}

\par \vspace{0.5cm}
Now we are in the position to state the first heredity result. 

\begin{theorem}\label{homo}
Conjecture \ref{conj} is true if $S$ is homoplasy-free.
\end{theorem}

\begin{figure}[ht]      \centering\vspace{0.5cm} 
    \includegraphics[width=15cm]{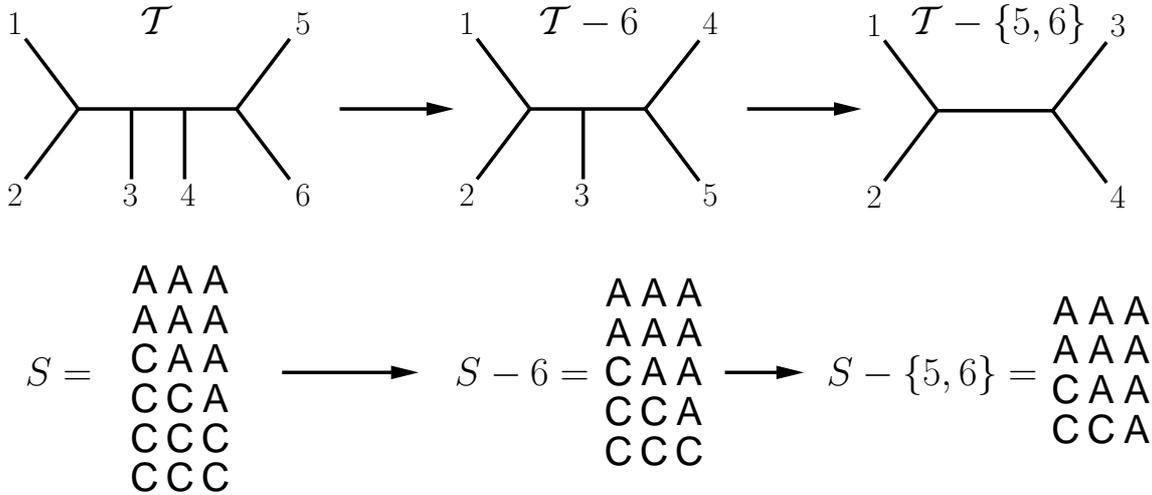} 
    \caption {Illustration of Theorem \ref{homo}. Alignment $S$ is homoplasy-free on tree $\T$, so all subalignments are homoplasy-free on the corresponding subtrees.}
   \label{homoplasyfreecase}
  \end{figure}

\begin{proof} 

Let $S$ be a homoplasy-free alignment with MP tree $\T$ and taxon set $X=\{1,\ldots,m\}$. Then, by Lemma \ref{homosubcharacter}, for any taxon $k \in X$ we conclude that the restriction $S-k$ of $S$ on $X-k$ is homoplasy-free on the corresponding restriction $\T-k$ of $\T$. As explained in Section \ref{notation}, homoplasy-free alignments are parsimoniously best possible, i.e. because $S-k$ is homoplasy-free on $\T-k$, $T-k$ is an MP tree for $S-k$. We repeat this argument to derive the desired sequence of MP trees from $m-1$ taxa down to 4 taxa. This completes the proof.
\end{proof}
An example for heredity of homoplasy-free alignments is depicted in Figure \ref{homoplasyfreecase}.

In order to investigate Conjecture \ref{conj} for general alignments, we now describe the idea underlying the following results. \begin{idea} If for $m$ taxa there exist $p$ distinct characters (or, more precisely, character patterns, cf. Section \ref{notation}) $f_1,\ldots, f_p$, the parsimony score of an alignment $S$ on an $m$-taxa tree $\T$ can be expressed as $\sum\limits_{i=1}^{p}x_i l_{\T}(f_i)$, where $x_i$ denotes the number of times the character $f_i$ occurs in $S$ (note that this implies  $|S| = \sum\limits_{i=1}^{p}x_i$). So the fact that a tree $\T$ is parsimoniously better than another tree $\hat{\T}$ concerning some alignment $S$ can be expressed in terms of the inequality $\sum\limits_{i=1}^{p}x_i l_{\T}(f_i) <\sum\limits_{i=1}^{p}x_i l_{\hat{\T}}(f_i).$ The same can be done for subalignments and the corresponding subtrees, so that altogether Conjecture \ref{conj} leads to a system of inequalities that need to be fulfilled by a potential counterexample. Such systems can then be tackled with the help of computer algebra systems. \end{idea}

We now use the idea explained above to prove the following statement on five taxa.

\begin{figure}[ht]      \centering\vspace{0.5cm} 
    \includegraphics[width=5cm]{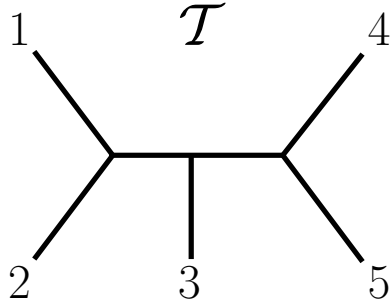} 
    \caption {Tree $\T=((1,2),3,(4,5))$, which is an MP tree for some given alignment $S$.}
   \label{5taxa_pic}
  \end{figure}

\begin{figure}[ht]      \centering\vspace{0.5cm} 
    \includegraphics[width=4cm]{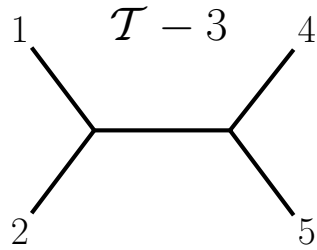} 
    \caption {If $\T=((1,2),3,(4,5))$ is an MP tree for some given alignment $S$, tree $\T-3=((1,2)(4,5))$, is an MP tree for $S-3$.}
   \label{5taxa_pic_4taxasub}
  \end{figure}

\begin{theorem}\label{5taxa}
Conjecture \ref{conj} is true in the case where $|X|=m=5$ and $S=f_1 \ldots f_n$ is binary, i.e. $f_1,\ldots, f_n$ are 2-state characters. In particular, if a tree $\T=((1,2),3,(4,5))$ as depicted in Figure \ref{5taxa_pic} is an MP tree for such an alignment $S$, then the tree $\T-3 = ((1,2)(4,5))$ as depicted in Figure \ref{5taxa_pic_4taxasub} is an MP tree for the alignment $S-3$, which results from $S$ when taxon $3$ is deleted.
\end{theorem}

\begin{proof} Let $f_1:=AACCC$, $f_2:=ACACC$, $f_3:=ACCAC$, $f_4:=ACCCA$, $f_5:=ACCAA$, $f_6:=ACACA$, $f_7:=ACAAC$, $f_8:=AACCA$, $f_9:=AACAC$, $f_{10}:=AAACC$ be the ten parsimoniously informative characters on five taxa. Let $S$ be a binary alignment on five taxa. Without loss of generality, we assume that tree $\T=((1,2),3,(4,5))$ as depicted in Figure \ref{5taxa_pic} is most parsimonious for $S$ (otherwise we re-label the leaves). This particularly implies that \begin{equation} \label{5taxa_T_better} l_{\T}(S) \leq l_{\hat{\T}}(S) \mbox{ as well as } l_{\T}(S) \leq l_{\tilde{\T}}(S), \end{equation} where $\hat{\T}=((1,4),3,(2,5))$ and $\tilde{\T}=((1,5),3,(2,4))$ are the trees depicted in Figure \ref{5taxa_pic_nonMP}. Note that we may ignore non-informative characters as they have the same score on all trees. Therefore, we can think of $S$ as a combination of characters $f_1,\ldots, f_{10}$, which occur $x_1,\ldots,x_{10}$ times in $S$, respectively. We then rewrite Inequality \ref{5taxa_T_better} as follows:
\begin{equation} \label{5taxa_T_better_b}  \sum\limits_{i=1}^{10}x_i l_{\T}(f_i) \leq \sum\limits_{i=1}^{10}x_i l_{\hat{\T}}(f_i) \mbox{ and } \sum\limits_{i=1}^{10}x_i l_{\T}(f_i) \leq \sum\limits_{i=1}^{10}x_i l_{\tilde{\T}}(f_i).\end{equation}
Calculating the parsimony scores for $f_1,\ldots,f_{10}$ on trees $\T$, $\hat{\T}$ and $\tilde{\T}$, respectively, we get $l_{\T}(f_1)=l_{\T}(f_{10})=l_{\hat{\T}}(f_3)=l_{\hat{\T}}(f_{7})=l_{\tilde{\T}}(f_4)=l_{\tilde{\T}}(f_{6})=1$, and all other parsimony scores are equal to 2. We now rewrite Inequality \ref{5taxa_T_better_b} using these scores: \par\vspace{-1.5cm}\begin{flushleft} \begin{eqnarray}\nonumber\mbox{\scriptsize $x_1+2x_2+2x_3+2x_4+2x_5+2x_6+2x_7+2x_8+2x_9+x_{10}$ }\hspace{-0.3cm}&\mbox{\scriptsize$\leq$} &  \hspace{-0.3cm} \mbox{\scriptsize $2x_1+2x_2+x_3+2x_4+2x_5+2x_6+x_7+2x_8+2x_9+2x_{10} $} \\ \label{conflict1}\Leftrightarrow x_3+x_7  & \leq & x_1+x_{10}  \\ & \nonumber\mbox{ and }& \\  
\nonumber\mbox{\scriptsize $x_1+2x_2+2x_3+2x_4+2x_5+2x_6+2x_7+2x_8+2x_9+x_{10}$ }\hspace{-0.3cm}&\mbox{\scriptsize$\leq$} &  \hspace{-0.3cm} \mbox{\scriptsize $2x_1+2x_2+2x_3+x_4+2x_5+x_6+2x_7+2x_8+2x_9+2x_{10} $} \\ \label{conflict2}\Leftrightarrow x_4+x_6  & \leq & x_1+x_{10} \end{eqnarray}\end{flushleft}

 \begin{figure}[ht]      \centering\vspace{0.5cm} 
    \includegraphics[width=12cm]{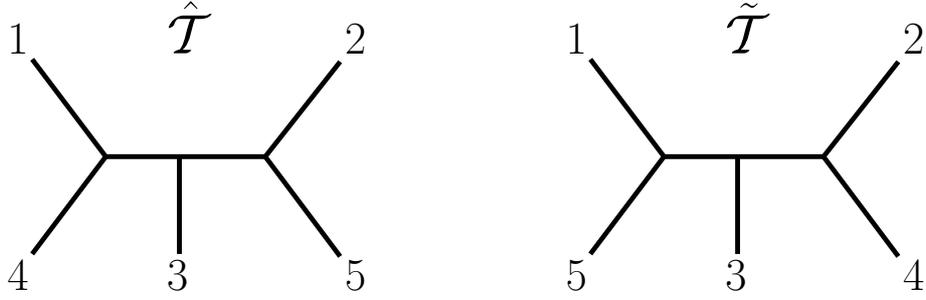} 
    \caption {If tree $\T$ depicted in Figure \ref{5taxa_pic} is an MP tree for an alignment $S$, the trees $\hat{\T}=((1,4),3,(2,5))$ and $\tilde{\T}=((1,5),3,(2,4))$ cannot have better parsimony scores for $S$ than $\T$, which leads to Inequality (\ref{5taxa_T_better}).}
   \label{5taxa_pic_nonMP}
  \end{figure}

Now we assume that the subtree $\T-3=((12),(45))$ of $\T$ is not most parsimonious. This implies that at least one of the two alternative trees, namely $\hat{\T}-3=((14),(25))$ or $\tilde{\T}-3=((15),(24))$, must be strictly better than $\T-3$ in the sense of parsimony. Using the above argument, we get \begin{equation} \label{4taxa_T_worse} l_{\T-3}(S-3) > l_{\hat{\T}-3}(S-3) \mbox{ or } l_{\T-3}(S-3) > l_{\tilde{\T}-3}(S-3). \end{equation} Calculating the parsimony scores of $f_1-3,\ldots,f_{10}-3$ on trees $\T-3$, $\hat{\T}-3$ and $\tilde{\T}-3$, respectively, we get $l_{\T-3}(f_3-3)=l_{\T-3}(f_{4}-3)=l_{\T-3}(f_{6}-3)=l_{\T-3}(f_{7}-3)=$\\$l_{\hat{\T}-3}(f_1-3)=l_{\hat{\T}-3}(f_{4}-3)=l_{\hat{\T}-3}(f_6-3)=l_{\hat{\T}-3}(f_{10}-3)=l_{\tilde{\T}-3}(f_1-3)=l_{\tilde{\T}-3}(f_{3}-3)=l_{\tilde{\T}-3}(f_7-3)=l_{\tilde{\T}-3}(f_{10}-3)=2$, and all other parsimony scores are equal to 1. We now rewrite Inequality \ref{4taxa_T_worse} using these scores: \par\vspace{-1.5cm}\begin{flushleft} \begin{eqnarray}\nonumber\mbox{\scriptsize $x_1+x_2+2x_3+2x_4+x_5+2x_6+2x_7+x_8+x_9+x_{10}$ }\hspace{-0.3cm}&\mbox{\scriptsize$>$} &  \hspace{-0.3cm} \mbox{\scriptsize $2x_1+x_2+x_3+2x_4+x_5+2x_6+x_7+x_8+x_9+2x_{10} $} \\ \label{conflict1reached} \Leftrightarrow x_3+x_7  & > & x_1+x_{10}  \\ & \nonumber\mbox{ or }& \\  
\nonumber\mbox{\scriptsize $x_1+x_2+2x_3+2x_4+x_5+2x_6+2x_7+x_8+x_9+x_{10}$ }\hspace{-0.3cm}&\mbox{\scriptsize$>$} &  \hspace{-0.3cm} \mbox{\scriptsize $2x_1+x_2+2x_3+x_4+x_5+x_6+2x_7+x_8+x_9+2x_{10} $} \\ \label{conflict2reached}\Leftrightarrow x_4+x_6  & > & x_1+x_{10} \end{eqnarray}\end{flushleft}

As either Inequality (\ref{conflict1reached}) or (\ref{conflict2reached}) must hold, this contradicts either (\ref{conflict1}) or (\ref{conflict2}). Therefore, $\T-3$ is an MP tree for $S-3$. 
\end{proof}

Next we show that the result presented in Theorem \ref{5taxa} cannot be generalized to $r$-state characters for $r>2$. In fact, not only is it possible that the particular subtree $((12),(45))$ of a most parsimonious tree $((12),3,(45))$ is not most parsimonious for the corresponding subalignment. It is even possible that the most parsimonious tree does not have any most parsimonious 4-taxa subtree at all. 

\begin{proposition}\label{5taxabadcase} Conjecture \ref{conj} is not generally true for multistate characters, even if the tree under consideration is the only MP tree.\end{proposition} 

\begin{proof}
We construct an explicit example employing three character states. Let $f_1,\ldots,f_{10}$ be defined as in the proof of Theorem \ref{5taxa}. Additionally, we define $f_{11}:=AACCT$, $f_{12}:=AACTC$, $f_{13}:=AATCC$, $f_{14}:=ACACT$, $f_{15}:=ACATC$, $f_{16}:=ATACC$, $f_{17}:=ACCAT$, $f_{18}:=ACTAC$, $f_{19}:=ATCAC$, $f_{20}:=ACCTA$, $f_{21}:=ACTCA$, $f_{22}:=ATCCA$, $f_{23}:=TAACC$, $f_{24}:=TACAC$ and $f_{25}:=TACCA$. Note that there is no parsimoniously informative 5-taxa character employing more than three states, so $f_1,\ldots,f_{25}$ is the complete list of parsimoniously informative characters on five taxa. Now if a tree $\T=((12),3,(45))$ shall be the unique MP tree for an alignment $S$ and none of the 4-taxa subtrees of $\T$, i.e. $\T-1=((23),(45))$, $\T-2=((13),(45))$, $\T-3=((12),(45))$, $\T-4=((12),(35))$ and $\T-5=((12),(34))$, shall be most parsimonious for the corresponding subalignments of $S$, this can be expressed with the help of the following system of inequalities (as in the proof of Theorem \ref{5taxa} we can ignore non-informative characters without loss of generality):
\par\vspace{-1.5cm}\begin{flushleft}
\begin{eqnarray*}  \mbox{\scriptsize for all $\hat{\T}\neq \T$ we have  }  \sum\limits_{i=1}^{25}x_i l_{\T}(f_i) &\leq &\sum\limits_{i=1}^{25}x_i l_{\hat{\T}}(f_i) \\ 
\mbox{ \parbox{5cm}{\center \scriptsize \par\vspace{-0.4cm}and for each $j = 1,\ldots, 5$ there \\ exists some tree $\tilde {\T}^j\neq \T$ such that} }\sum\limits_{i=1}^{25}x_i l_{\T-j}(f_i-j) & > &\sum\limits_{i=1}^{25}x_i l_{\tilde{\T}^j-j}(f_i-j).
\end{eqnarray*}\end{flushleft}

Using a computer algebra system, we find that one possible solution is $x_4=1$, $x_6=2$, $x_{11}=1$, $x_{13}=2$, $x_{23}=1$ and all other $x_i=0$. This gives the following alignment:

$$ S := 
\left\{\begin{array}{ccccccc} \vspace{-0.5em}
A & A &A & A & A & A & A \\ \vspace{-0.5em}
C & C &C & A & A & A & C \\ \vspace{-0.5em}
C & A &A & C & C & C & C \\ \vspace{-0.5em}
C & C &C & C & T & T & T \\ \vspace{-0.5em}
A & A &A & T & T & T & T \\ \vspace{-0.5em}
\end{array}\right.
$$

So $S$ is an alignment with unique MP tree $\T=((12),3,(45))$ and no most parsimonious 4-taxa subtree, which can be verified by examining all five distinct characters employed by $S$ on all 15 trees on five taxa and their corresponding 4-taxa subtrees.
\end{proof}

So for five taxa, the question whether or not Conjecture \ref{conj} holds depends on the number of character states the alignment employs: for two character states it holds, whereas it fails for three or more states. Next we show that this distinction cannot be generalized to more than five taxa. In fact, we use our approach of solving inequality systems in order to generate an alignment $S$ on six taxa with unique MP tree $\T=(((12),3),(4,(5,6)))$, which has neither a most parsimonious 5-taxa subtree nor a most parsimonious 4-taxa subtree for the corresponding subalignments of $S$.

\begin{proposition}\label{6taxabadcase} Conjecture \ref{conj} is not generally true for more than five taxa, even if only binary characters are employed and the tree under consideration is the only MP tree. In fact, a unique MP tree might not have any (non-trivial) most parsimonious subtree at all. \end{proposition} 

\begin{proof}
Consider all informative binary characters on six taxa: $f_{1}= AACCCC$, $f_{2}= ACACCC$, $f_{3}= ACCACC$, $f_{4}= ACCCAC$, $f_{5}= ACCCCA$, $f_{6}= ACCAAA$, $f_{7}= ACACAA$, $f_{8}= ACAACA$, $f_{9}= ACAAAC$, $f_{10}= AACCAA$, $f_{11}= AACACA$, $f_{12}= AACAAC$, $f_{13}= AAACCA$, $f_{14}= AAACAC$, $f_{15}= AAAACC$, $f_{16}= AAACCC$, $f_{17}= AACACC$, $f_{18}= AACCAC$, $f_{19}= AACCCA$, $f_{20}=ACAACC$, $f_{21}= ACACAC$, $f_{22}= ACACCA$, $f_{23}= ACCAAC$, $f_{24}= ACCACA$ and $f_{25}= ACCCAA$. 
Now we construct an example analogously to the construction shown in the proof of Proposition \ref{5taxabadcase} and find that the alignment $S$ employing two copies of $f_3$, five copies of $f_4$, four copies of $f_5$, one copy of $f_7$, nine copies of $f_8$, six copies of $f_9$, eleven copies of $f_{11}$, nine copies of $f_{12}$, three copies of $f_{13}$, two copies of $f_{14}$, seven copies of $f_{16}$, one copy of $f_{19}$, four copies of $f_{20}$ and six copies of $f_{25}$ has the desired properties. This alignment is depicted in Figure \ref{strongcounterex}. It has a unique MP tree, namely $\T = (((1,2),3),(4,(5,6)))$ as depicted in Figure \ref{6taxauniqueMP}. The 5-taxa and 4-taxa MP trees for $S$ are depicted in Figures \ref{mp_sub5} and \ref{mp_sub4}, respectively. If the reader wishes to verify that these results are correct, we strongly recommend the program `Penny' from the free Phylip-package \citep{phylip}, which is able to run an exhaustive search through the tree space for binary character sequences.
\end{proof}

 \begin{figure}[ht]      \centering\vspace{0.5cm} 
 $
\begin{array}{c} \mathtt{
AAAAAAAAAAA A AAAAAAAAA AAAAAAAAAAAAAAAAAAAAAAAAAA AAAAAAAAAAAAAAAAAAAAAAA}\\
\mathtt{CCCCCCCCCCCCCCCCCCCCCCCCCCCAAAAAAAAAAAAAAAAAAAAAAAAAAAAAAAAACCCCCCCCCC}\\
\mathtt{CCCCCCCCCCCAAAAAAAAAAAAAAAACCCCCCCCCCCCCCCCCCCCAAAAAAAAAAAACAAAACCCCCC}\\
\mathtt{AACCCCCCCCCCAAAAAAAAAAAAAAAAAAAAAAAAAAAAAAAAAAACCCCCCCCCCCCCAAAACCCCCC}\\
\mathtt{CCAAAAACCCCACCCCCCCCCAAAAAACCCCCCCCCCCAAAAAAAAACCCAACCCCCCCCCCCCAAAAAA}\\
\mathtt{CCCCCCCAAAAAAAAAAAAAACCCCCCAAAAAAAAAAACCCCCCCCCAAACCCCCCCCCACCCCAAAAAA}
\end{array}$
   \caption {Alignment $S$ as defined in the proof of Proposition \ref{6taxabadcase} has a unique MP tree $\T$ (cf. Figure \ref{6taxauniqueMP}), which has no MP subtrees.}   \label{strongcounterex}
     \end{figure}

 \begin{figure}[ht]      \centering\vspace{0.5cm} 
    \includegraphics[width=6cm]{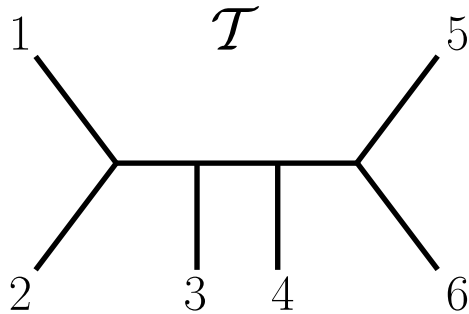} 
    \caption {Tree $\T=(((1,2),3),(4,(5,6)))$ is the unique MP tree for $S$ shown in Figure \ref{strongcounterex} but has no most parsimonious subtrees.}
   \label{6taxauniqueMP}
  \end{figure}

 \begin{figure}[ht]      \centering\vspace{0.5cm} 
    \includegraphics[width=15cm]{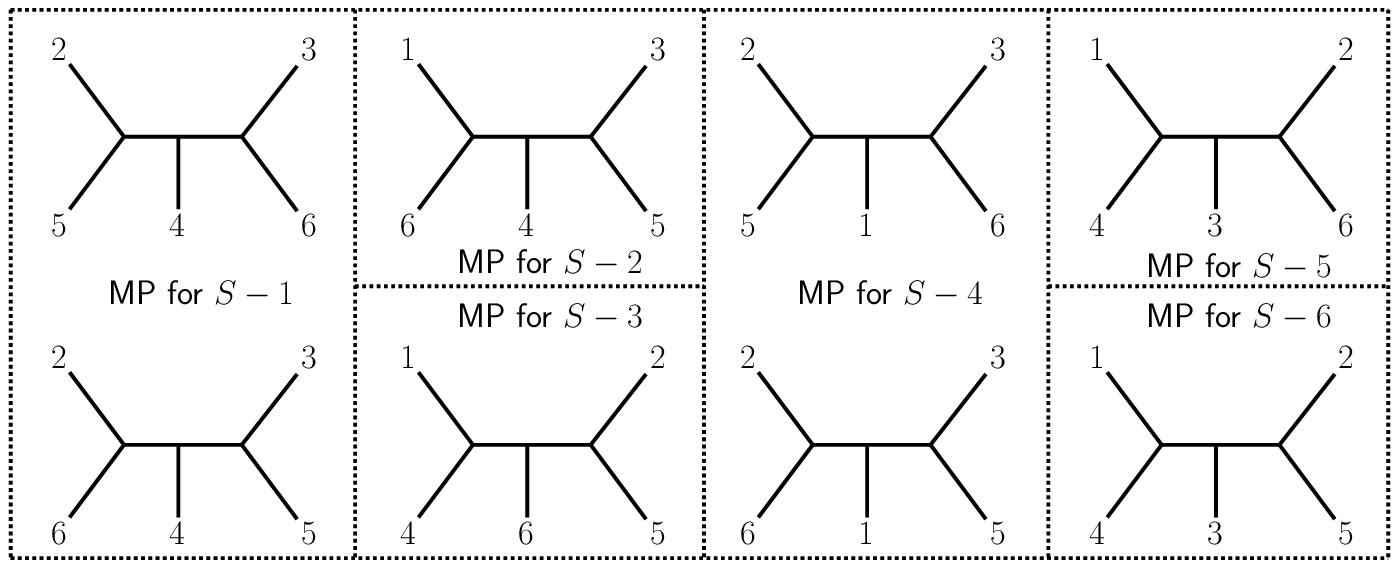} 
    \caption {Illustrations of all 5-taxa MP trees for the corresponding subalignments of alignment $S$ as defined in the proof of Proposition \ref{6taxabadcase}. None of these trees is a subtree of $\T$ shown in Figure \ref{6taxauniqueMP}, which is the unique MP tree of $S$.}
   \label{mp_sub5}
  \end{figure}

\begin{figure}[ht]      \centering\vspace{0.5cm} 
    \includegraphics[width=15cm]{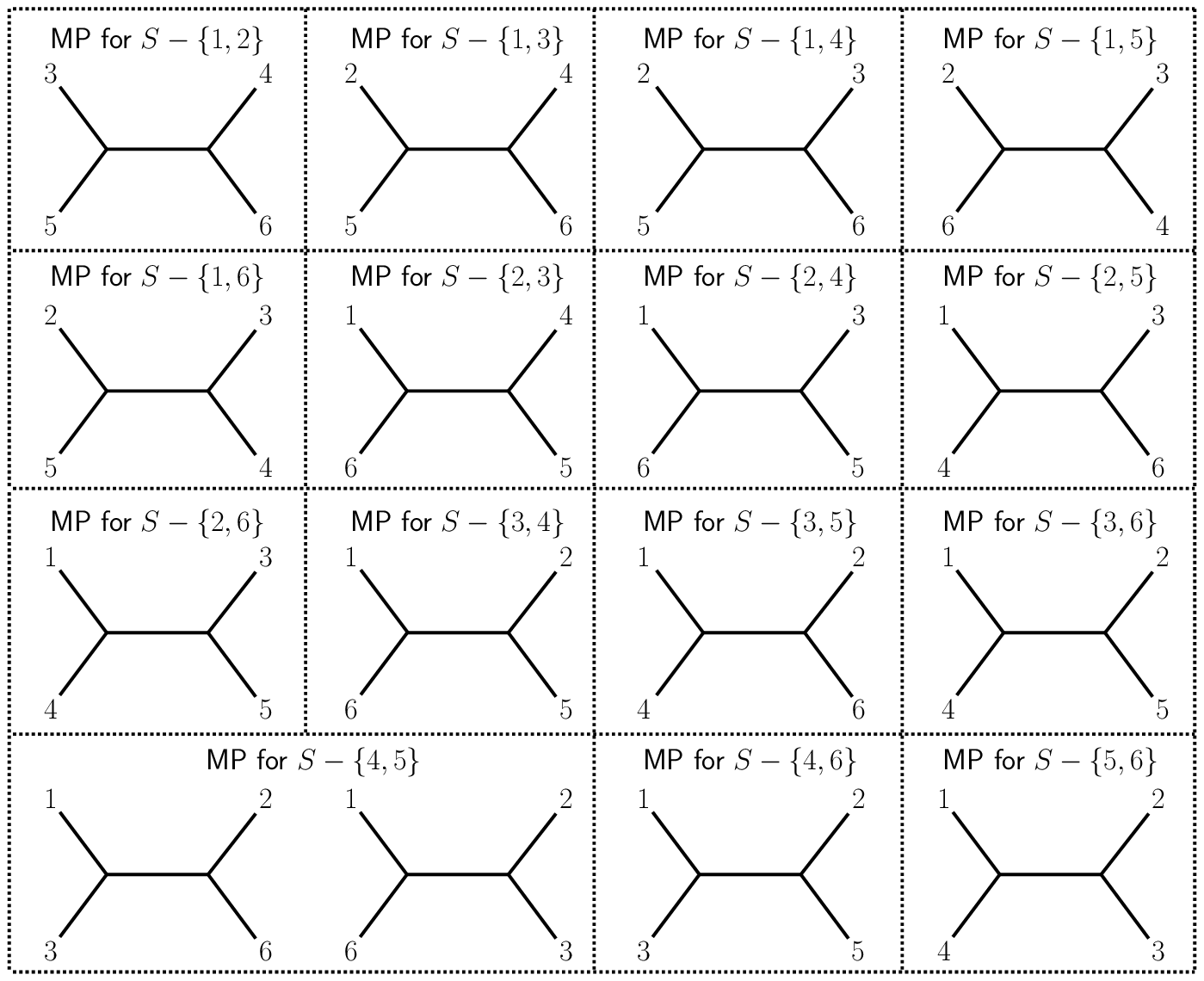} 
    \caption {Illustrations of all 4-taxa MP trees for the corresponding subalignments of alignment $S$ as defined in the proof of Proposition \ref{6taxabadcase}. None of these trees is a subtree of $\T$ shown in Figure \ref{6taxauniqueMP}, which is the unique MP tree of $S$.}
   \label{mp_sub4}
  \end{figure}

Note that the example presented in the proof of Theorem \ref{6taxabadcase} concerns tree $\T=(((1,2),3),(4,(5,6)))$, analogous examples can be constructed for the other tree shape on six taxa, namely $(((1,2),(3,4),(5,6)))$ (example not shown). We conclude that in general, MP trees for an alignment do not have to be related to MP trees on subalignments. This surprising result shows once again that MP, while being a simple combinatorial algorithm, is more complicated than one might intuitively think. As explained above, the existence of such instances is not immediately clear because of the NP-hardness of finding the set of most parsimonious trees in the tree space. It rather adds another complicated aspect to an already hard problem.

\subsection{Heredity Part II: Constructing large MP trees from smaller ones} \label{benny}

In the previous section, we showed that Maximum Parsimony trees are not in general hereditary in the sense of allowing for the inference of smaller MP trees by known larger ones. In the present section, we approach a different aspect of heredity: Given an alignment, is it possible to combine small compatible MP trees, e.g. quartets, to derive an MP tree for the entire set of taxa? Intuitively, one might think that it is quite likely that this is true, as the asumption of compatibility of the MP-quartets is a strong condition. Moreover, one might think that if additionally the MP-quartets are all unique, which is another strong condition, it is even more likely for such a statement to hold. However, in this section we present a counterexample which shows that even under these seemingly ideal conditions it may be impossible to infer large MP trees from the smaller ones.

\begin{proposition} \label{bennyprop} If for an alignment $S$ on the taxa set $X$ all most parsimonious quartet trees (for taxa sets $\{x_1,x_2,x_3,x_4\}\subseteq X$) on the corresponding subalignments of $S$ are compatible with an $X$-tree $\hat{\T}$, this tree $\hat{\T}$ does not need to be an MP tree for $S$. This is even true if the MP-quartets and the MP tree for $S$ are all unique.
\end{proposition}

\begin{figure}[ht]      \centering\vspace{0.5cm} 
    \includegraphics[width=15cm]{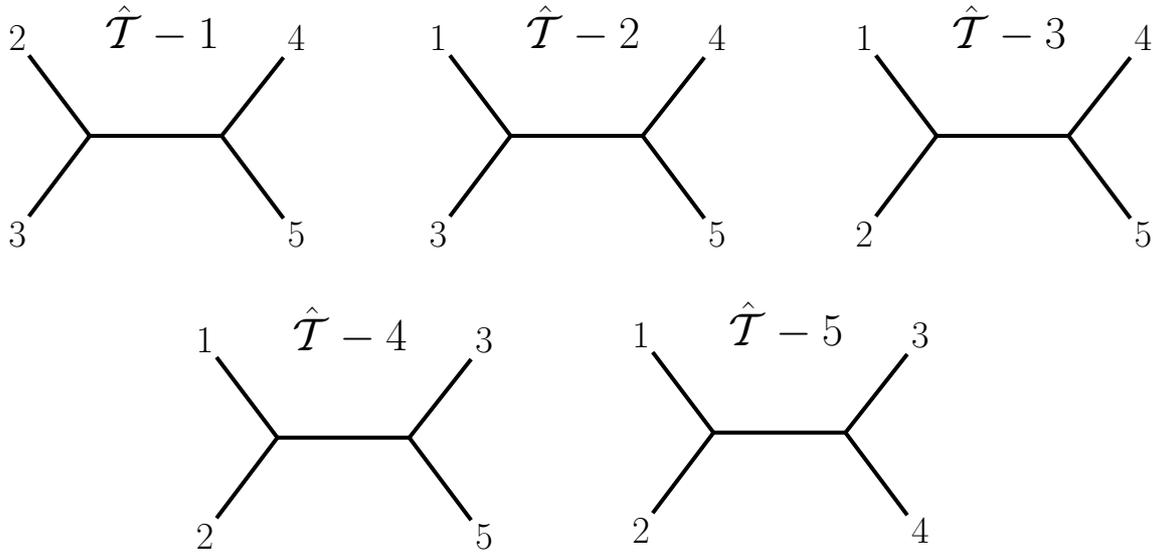} 
    \caption {Illustration of the unique most parsimonious quartet trees of $S$ as defined in the proof of Proposition \ref{bennyprop}.}
   \label{benny_fig1}
  \end{figure}

\begin{figure}[ht]      \centering\vspace{0.5cm} 
     \includegraphics[width=6cm]{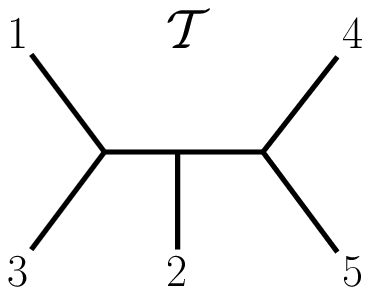} \hspace{2cm}
\includegraphics[width=6cm]{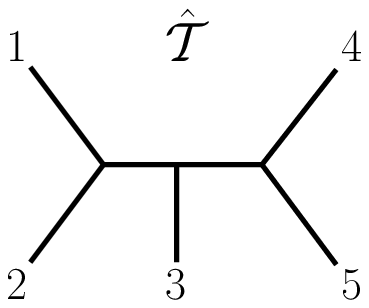} 
    \caption {Tree $\T$ is the unique MP tree of $S$ as defined in the proof of Proposition \ref{bennyprop}, but tree $\hat{\T}=((1,2),3,(4,5))$ is the only tree that is compatible with all unique MP quartet trees.}
   \label{benny_fig2}
  \end{figure}

\begin{proof} We prove the proposition by providing an explicit counterexample. Consider the following binary alignment:
$$ S := 
\left\{\begin{array}{ccccccccccccccc} \vspace{-0.5em}
A & A &A & A & A & A & A &A & A & A &A & A &A & A & A  \\ \vspace{-0.5em}
A & A &C & C & C&A & A &A & A & A &A & A &A & A & A   \\ \vspace{-0.5em}
C & C &A & A & A & C & C &C & C & C&C& A &A & A & A  \\ \vspace{-0.5em}
C & C &C & C & C&C & C &C &A & A & A&C & C &C &C  \\ \vspace{-0.5em}
C & C &C & C & C  &A & A & A&C & C &C &C&C & C &C \\ \vspace{-0.5em}
\end{array}\right.
$$
$S$ consists of two copies of $f_1$, three copies of $f_2$, three copies of $f_8$, three copies of $f_9$ and four copies of $f_{10}$, where $f_1,\ldots,f_{10}$ are defined as in the proof of Theorem \ref{5taxa}. Using an exhaustive search through the space of all 15 trees on five taxa, which for binary alignments is provided e.g. by the `Penny' program of the Phylip package \citep{phylip}, we find that $S$ has the unique MP tree $\T = ((1,3),2,(4,5))$ depicted in Figure \ref{benny_fig2}. Moreover, subalignment $S-1$ has the unique MP quartet tree $\hat{\T}-1:= ((2,3),(4,5))$, subalignment $S-2$ has the unique MP quartet tree $\hat{\T}-2:= ((1,3),(4,5))$, subalignment $S-3$ has the unique MP quartet tree $\hat{\T}-3:= ((1,2),(4,5))$, subalignment $S-4$ has the unique MP quartet tree $\hat{\T}-4:= ((1,2),(3,5))$ and subalignment $S-5$ has the unique MP quartet tree $\hat{\T}-5:= ((1,2),(3,4))$. All these trees are depicted in Figure \ref{benny_fig1}. Note that trees $\hat{\T}-1$, $\hat{\T}-2$, $\hat{\T}-3$, $\hat{\T}-4$  and $\hat{\T}-5$ are all compatible with tree $\hat{\T}:=((1,2),3,(4,5))$ depicted in Figure \ref{benny_fig2}, whereas $\hat{\T}-4$  and $\hat{\T}-5$ are incompatible with $\T$. So the unique and compatible MP quartets cannot be combined to give the unique MP tree for the whole alignment. This completes the proof.
\end{proof}

While Section \ref{hereditary1} shows that in general large MP trees cannot be used to infer smaller MP trees on subsets of the taxon set, the above example shows that the opposite is also impossible, even under strong compatibility conditions. So MP is a phylogenetic tree inference method that may find that unique `best' trees are unrelated to `best' trees on subsets or supersets of the taxa under consideration. As this is somewhat counterintuitive, naturally the question arises whether this problem only occurs with MP or also affects other methods. In the next section, we will generalize our results to Maximum Likelihood (ML) under a frequently used nucleotide substitution model.

\subsection{Heredity Part III: Impacts of the parsimony results on Maximum Likelihood}\label{sec_ML}

In the following, we will examine the impacts of the results presented in Sections \ref{hereditary1} and \ref{benny} on Maximum Likelihood (ML) under the so-called $N_r$-model introduced in \citep{neyman_1971} and explained in Section \ref{notation}. The $N_4$-model is also known as the Jukes Cantor model \citep{jukes_cantor_1969}. It may be biologically justified to assume that the substitution probabilities on an edge of the tree are the same for each character in an alignment, particularly if the characters are close to one another in the alignment. In this case, we say these characters evolved under a {\it common mechanism}. If, on the contrary, different mechanisms are allowed to operate at each site, we say there is {\it no common mechanism}. For both cases, Tuffley and Steel presented results that closely link Maximum Likelihood with Maximum Parsimony \citep{tuffley_steel_1997}.

\begin{theorem} [Equivalence of MP and ML under no common mechanism, Theorem 5 of \citep{tuffley_steel_1997}] \label{tuffley5}Maximum parsimony and maximum likelihood with no common mechanism are equivalent in the sense that both choose the same tree or trees.
\end{theorem}

\begin{theorem}[Agreement of ML with MP under the $N_r$-model, Theorem 7 of \citep{tuffley_steel_1997}]  \label{tuffley7}For data containing enough constant characters, the maximum likelihood tree under the $N_r$-model is a maximum parsimony tree.
\end{theorem}

The consequences of these theorems in combination with Sections \ref{hereditary1} and \ref{benny} can be described as follows: if we assume a common mechanism, all examples provided in these sections immediately lead to analogous results for Maximum Likelihood by Theorem \ref{tuffley5}. If no common mechanism is assumed, the examples provided in these sections may need to be modified in the sense of adding constant characters. These extra characters do not change the MP tree as constant characters are non-informative, but they will make ML agree with MP according to Theorem \ref{tuffley7}. So in both cases, we derive alignments for which the ML tree is not hereditary. The following corollaries are therefore direct conclusions from the previous sections combined with Theorems \ref{tuffley5} and \ref{tuffley7}.

\begin{corollary} Let $S:= f_1f_2\ldots f_n$ be a sequence of characters (alignment) on the set $X$ of taxa, where $|X|=m\geq 5$, and let $\T$ be a Maximum Likelihood tree for $S$. Then, under the $N_r$-model (with or without assuming a common mechanism), there may be a subset $Y$ of $X$ of size at least four, such that $\T|_Y$ is {\it not} an ML tree for $S|_Y$ (where $S|_Y$ is the sequence $S$ of characters restricted to the taxa in $Y$ and $\T|_Y$ is the tree $\T$ restricted to the taxa in $Y$). In fact, a unique ML tree might not have any (non-trivial) most likely subtree at all.
\end{corollary} 

\begin{proof} This result is a direct conclusion of Theorems \ref{tuffley5} and \ref{tuffley7}, respectively, to the examples (if required filled up with sufficiently many constant characters) constructed in the proofs of Propositions \ref{5taxabadcase} and \ref{6taxabadcase}.
\end{proof}

\begin{corollary} If for an alignment $S$ on the taxa set $X$ all ML quartet trees (for taxa sets $\{x_1,x_2,x_3,x_4\}\subseteq X$) on the corresponding subalignments of $S$ are compatible with an $X$-tree $\hat{\T}$, this tree $\hat{\T}$ does not need to be an ML tree for $S$. This is even true if the ML-quartets and the ML tree for $S$ are all unique.
\end{corollary}

\begin{proof} This result is a direct conclusion of Theorems \ref{tuffley5} and \ref{tuffley7}, respectively, to the example (if required filled up with sufficiently many constant characters) constructed in the proof of Proposition \ref{bennyprop}.
\end{proof}

\section{Discussion}\label{discussion}

In this paper, we presented various examples of non-hereditary Maximum Parsimony and Maximum Likelihood trees together with an idea of how to construct them as solutions to systems of inequalities. The results show that there are alignments for which the `best' tree with respect to one of these phylogenetic tree inference methods does not have to be related to the `best' tree on fewer taxa. Also, even if a tree is constructed from uniquely `best' and compatible quartet trees, it might not coincide with the `best' tree on all taxa. On the one hand, these facts might help to understand why tree reconstruction for MP and ML is hard (\citep{foulds_graham_1982}, \citep{roch_2006} ,\citep{chor_tuller_2006}). On the other hand, the result is surprising and gives rise to new questions, e.g. whether or not one should include outgroups when inferring trees, as these might change the tree topology of the optimal tree. Naturally, it would also be interesting to know whether similar heredity problems occur with other methods, such as e.g. distance methods or ML under more complicated models of nucleotide substitution. We conjecture that these methods are also not hereditary in the sense described in this paper.

\par\vspace{1.5cm}
\begin{acknowledgement} \upshape Some of the results underlying this paper were achieved during a summer studentship funded by the New Zealand Marsden Fund and hosted by the Allan Wilson Centre for Molecular Ecology and Evolution. I also wish to thank Mike Steel and Arndt von Haeseler for helpful discussions on the topic and Benny Chor for asking a question that led to the results presented in Section \ref{benny}. \end{acknowledgement}

\newpage
\bibliography{fischer_bibfile}

\begin{thebibliography}{12}
\providecommand{\natexlab}[1]{#1}
\providecommand{\url}[1]{\texttt{#1}}
\expandafter\ifx\csname urlstyle\endcsname\relax
  \providecommand{\doi}[1]{doi: #1}\else
  \providecommand{\doi}{doi: \begingroup \urlstyle{rm}\Url}\fi

\bibitem[Chor and Tuller(2006)]{chor_tuller_2006}
B.~Chor and T.~Tuller.
\newblock Finding a maximum likelihood tree is hard.
\newblock \emph{J. of the ACM}, 53:\penalty0 722 -- 744, 2006.

\bibitem[Felsenstein(1978)]{felsenstein_1978}
J.~Felsenstein.
\newblock Cases in which parsimony or compatibility will be positively
  misleading.
\newblock \emph{Syst. Zool.}, 27:\penalty0 401--410, 1978.

\bibitem[Felsenstein(2004)]{felsenstein_2004}
J.~Felsenstein.
\newblock \emph{Inferring phylogenies.}
\newblock Sinauer Associates, Massachusetts, 2004.

\bibitem[Felsenstein(2005)]{phylip}
J.~Felsenstein.
\newblock Phylip (phylogeny inference package version 3.6.
\newblock \emph{Distributed by the author. Department of Genome Sciences,
  University of Washington, Seattle.}, 2005.

\bibitem[Fitch(1971)]{fitch_1971}
W.~Fitch.
\newblock Toward defining the course of evolution: minimum change for a
  specific tree topology.
\newblock \emph{Syst. Zool.}, 20(4):\penalty0 406--416, 1971.

\bibitem[Foulds and Graham(1982)]{foulds_graham_1982}
L.R. Foulds and R.L. Graham.
\newblock The steiner problem in phylogeny is np-complete.
\newblock \emph{Adv. Appl. Math.}, 3:\penalty0 43--49, 1982.

\bibitem[Jukes and Cantor(1969)]{jukes_cantor_1969}
T.~Jukes and C.~Cantor.
\newblock Evolution of protein molecules.
\newblock \emph{In ``Mammalian Protein Metabolism", New York Academic Press},
  pages 21--132, 1969.

\bibitem[Neyman(1971)]{neyman_1971}
J.~Neyman.
\newblock Molecular studies of evolution: A source of novel statistical
  problems.
\newblock \emph{In ``Statistical Decision Theory and Related Topics", New York
  Academic Press}, pages 1--27, 1971.

\bibitem[Roch(2006)]{roch_2006}
S.~Roch.
\newblock A short proof that phylogenetic tree reconstruction by maximum
  likelihood is hard.
\newblock \emph{IEEE/ACM Trans. on Comp. Biol. and Bioinf.}, 3:\penalty0
  92--94, 2006.

\bibitem[Semple and Steel(2003)]{semple_steel_2003}
C.~Semple and M.~Steel.
\newblock \emph{Phylogenetics.}
\newblock Oxford University Press, 2003.

\bibitem[Tuffley and Steel(1997)]{tuffley_steel_1997}
C.~Tuffley and M.~Steel.
\newblock Links between maximum likelihood and maximum parsimony under a simple
  model of site substitution.
\newblock \emph{Bull. Math. Biol.}, 59:\penalty0 581--607, 1997.

\bibitem[Yang(2006)]{yang_2006}
Z.~Yang.
\newblock \emph{Computational Molecular Evolution.}
\newblock Oxford University Press, 2006.

\end{thebibliography}

\end{document}